\documentclass[times,doublespace]{qjrms3}

\usepackage[dvips,colorlinks,bookmarksopen,bookmarksnumbered,citecolor=red,urlcolor=red]{hyperref}

\usepackage{moreverb}

\begin{document}

\runningheads{M. Akhalkatsi, G. Gogoberidze}{Spectrum of infrasound
from supercell storms}

\title{Spectrum of infrasound radiation from supercell storms}

\author{M. Akhalkatsi\affil{a}, G. Gogoberidze\affil{a}\corrauth}

\address{Institute of theoretical physics, Ilia State University, 2a Kazbegi ave., 0160 Tbilisi, Georgia}

\corraddr{E-mail: gogober@geo.net.ge}

\begin{abstract}

We consider the generation of acoustic waves by turbulent convection
and perform spectral analysis of a monopole source of sound related
to the heat production by condensation of moisture.  A quantitative
explanation of the correlation between intensity of infrasound
generated by supercell storms and later tornado formation is given.
It is shown that low lifting condensation level (LCL) and high
values of convective available potential energy (CAPE), which are
known to favor significant tornadoes, also lead to a strong
enhancement of supercell's low frequency acoustic radiation.

\end{abstract}


\received{}
\revised{}
\accepted{}

\maketitle

\section{Introduction}
Recent observations of infrasound originating from regions of severe
weather  show that infrasound dominant frequency occurs in a
passband from 0,5 to 2,5 Hz, with peak frequencies between 0,5 and 1
Hz (e.g. Bedard, 2005; Bedard \emph {et al}., 2004a). Analyzing
acoustic radiation from severe thunderstorms, Bedard (2004a)
concluded that radiation of infrasound in this passband is not a
natural consequence of all severe weather. Infrasound of a tornadic
thunderstorm is much stronger than infrasound of a nonsevere weather
system and therefore, mesocyclones or tornadoes may be a primary
mechanism for infrasound production in this frequency range (Szoke
\emph {et al}., 2004).

Detection of infrasound appears to have significant potential for
improving tornado forecasting. The acoustic power radiated by strong
convective storm system could be as high as $10^7$ watts (Georges,
1988) and infrasound below 1 Hz can travel for distances of
thousands of kilometers from a source without significant
absorption.

Over the years, several potential sound generation mechanisms were
compared with measured characteristics of infrasound (Georges and
Greene, 1975; Georges, 1988; Bedard and Georges, 2000; Beasley \emph
{et al}., 1976). Such mechanisms include release of latent heat,
dipole radiators, boundary layer turbulence, lightning,
electrostatic sources and vortex sound (radial vibrations and the
co-rotation of suction vortices). Georges (1976) eliminated many
sources as likely candidates and concluded that vortex sound was the
most likely model. Bedard (2005) also found that the radial
vibration model (Abdullah, 1966) is most consistent with infrasonic
data. This model predicts that the fundamental frequency of radial
vibration will be inversely proportional to core radius. A radius of
about 200 m will produce a frequency of 1 Hz. Schecter \emph{et al}.
(2008) performed numerical simulation of the adiabatic generation of
infrasound by tornadoes and also simulated the infrasound radiated
from a single-cell non-tornadic thunderstorm in a shear-free
environment, using Regional Atmospheric Modeling System (RAMS). In
the latter simulation dominant infrasound in the $0.1-10~{\rm Hz}$
frequency band, so called "thunderstorm noise" appeared to radiate
from the vicinity of the melting level, where diabatic processes
involving hail were active. They found that 3D Rossby waves of a
tornado-like vortices can generate stronger $0.1-10~{\rm Hz}$
infrasound at or above the simulated non-tornadic thunderstorm noise
if maximum wind speed of the vortex exceeds a modest threshold.

Georges and Green (1975) noted that infrasound often precedes an
observed tornado by up to an hour. Many other case studies of
infrasound associated with tornadoes and tornadic storms (Bedard
\emph {et al}., 2004a; Bedard \emph {et al}., 2004b) are consistent
with this result. Bedard \emph {et al}. (2004b) analyzed all
significant infrasonic signals data collected in 2003 during the
continuous Infrasonic Network (ISNeT) operation. Summarizing
differences between predicted acoustic signal arrival times from
reported tornados and start times of infrasonic detection they
concluded that infrasound is usually produced substantially before
(0.5 - 1 hrs) reports of tornadoes and that there could be other
sound generation processes active not related to tornadic vortices.
Bedard \emph {et al}. (2004b) also indicated the difference between
the infrasonic bearing sectors and the vortex location and concluded
that, although the storm environment wind and temperature gradient
could be responsible for bearing deviations, there remains the
possibility that another storm feature could radiate infrasound from
another location within the storm (bedard, 2004).

Broad and smooth spectra of observed infrasound radiation indicates
that turbulence is a promising sources of the radiation. Acoustic
radiation from turbulent convection was studied by Akhalkatsi and
Gogoberidze (2009) taking into account effects of stratification,
temperature fluctuations and moisture of air, using Lighthill's
acoustic analogy. It was shown for typical parameters of strong
convective storms, infrasound radiation should be dominated by a
monopole acoustic source related to the moisture of the air. The
total power of this source is two orders higher than thermo-acoustic
and Lighthill's quadrupole radiation power, being of order
$10^7~{\rm watts}$, in qualitative agreement with observations of
strong convective storms (Bowman and Bedard, 1971; Bedard and
Georges, 2000; Georges and Greene, 1975; Georges, 1973).

This paper represents an extension of the study performed by
Akhalkatsi and Gogoberidze (2009). Namely, we extended earlier study
in two directions. Firstly, we perform detailed spectral analysis of
a monopole source related to heat production during the condensation
of moisture. Assuming homogeneous and stationary turbulence we
calculate the spectrum of acoustic radiation. Secondly, we perform
detailed analysis of the sound generated by a monopole source as
well as infrasound observations and present a qualitative
explanation of the observed high correlation between intensity of
low frequency infrasound signals from supercell storms and the
probability of later tornado formation. We show that acoustic power
of a monopole source related to the moisture of the air strongly
depends on the same parameters that are the most promising in
discriminating between nontornadic and tornadic supercells according
to the recent study of tornadogenesis (Markowsky and Richardson,
2008). In particular, low lifting condensation level (LCL) is known
to favor significant tornadoes (Rasmussen and Blanchard, 1998;
Thompson \emph {et al}., 2003). On the other hand, low LCL height
means that air in the updraft is saturated at lower heights and
consequently has a higher temperature at the level of saturation. We
show that the increase of temperature causes rapid enhancement of
monopole acoustic power related to heat production during
condensation of moisture.

Another widely used supercell and tornado forecasting parameter is
supercell environmental convective available potential energy
(CAPE). It is known that high values of CAPE (especially, occurring
closer to the surface) assist tornado formation (Weisman and Klemp,
1982; Rotunno and Klemp, 1982; Rasmussen, 2003; Rasmussen and
Blanchard, 1998; Thompson \emph {et al}., 2003). Furthermore, high
values of CAPE means high updraft velocity and therefore the
increased rms of turbulent velocities, which results in a strong
enhancement of total acoustic power.

The paper is organized as follows: General formalism is presented in
section 2. Spectral analysis of acoustic radiation is performed in
section 3. Correlation between intensity of infrasound radiation by
supercell storm and probability of tornado formation is discussed in
section 4. Conclusions are given in section 5.

\section{General formalism} \label{sec:2}
The dynamics of convective motion of moist air is governed by the
continuity, Euler, heat, humidity and ideal gas state equations:
\begin{equation}
\frac{D \rho}{D t} + \rho {\bf \nabla} \cdot {\bf v}=0,\label{eq:21}
\end{equation}
\begin{equation}
\rho\frac{D {\bf v}}{D t}=-{\bf \nabla}p- \rho {\bf
\nabla}\Phi,\label{eq:22}
\end{equation}
\begin{equation}
T\frac{D s}{D t}=-L_\nu \frac{D q}{D t},\label{eq:23}
\end{equation}
\begin{equation}
\rho=\frac{p}{R T}\frac{1}{1-q+q/\epsilon}=\frac{p}{R
T}\frac{1}{1+aq},\label{eq:25}
\end{equation}
where ${\bf v},~\rho$ and $p$ are velocity, density and pressure,
respectively;
 $D/Dt\equiv\partial/\partial t+{\bf v}\cdot
{\bf \nabla}$ is Lagrangian time derivative; $L_\nu$ is the latent
heat of condensation and $q$ is the humidity mixing ratio
\begin{equation} q\equiv \frac{\rho_\nu}{\rho}.\label{eq:26}
\end{equation}
Here $\rho_\nu$ is the mass of water vapor in a unit volume; $\Phi$
is gravitational potential energy per unit mass and ${\bf
\nabla}\Phi\equiv-{\bf g}$. $\epsilon \equiv m_\nu/m_d \approx
0.622$ is the ratio of molecular masses of water and air; $a=0.608$
and $R$ is the universal gas constant.

In the set of Eqs. (\ref{eq:21})-(\ref{eq:25}) diffusion and
viscosity effects are neglected due to the fact that they have a
negligible influence on low frequency acoustic wave generation and
propagation.

Using the standard procedure of Lighthill's acoustic analogy
(Goldstein 2002), manipulation of Eqs. (\ref{eq:21})-(\ref{eq:25})
yield the equation governing sound generation by turbulence for a
moist atmosphere in terms of the total enthalpy (Akhalkatsi and
Gogoberidze, 2009)
\begin{eqnarray}
&& \left( \frac{D}{Dt} \left(\frac{1}{c_s^2} \frac{D}{Dt} \right)
-\frac{{\bf \nabla}p \cdot {\bf \nabla}}{\rho c_s^2} -{\bf
\nabla}^2 \right)B= \nonumber\\
&& S_L + S_T + S_q + S_m + S_\gamma, \label{eq:215}
\end{eqnarray}
where $\gamma \equiv c_p/c_v$ is the ratio of specific heats,
\begin{equation}
c_s \equiv \left( \frac{\partial p}{\partial \rho}
\right)_{s,q}^{1/2} \label{eq:212}
\end{equation}
is the sound velocity and enthalpy $B$ is determined by the
generalized Bernoulli theorem
\begin{equation}
B\equiv E+\frac{p}{\rho} +\frac{v^2}{2}+\Phi, \label{eq:27}
\end{equation}
where $E$ is internal energy. $B$ is constant in the steady
irrotational flow and at large distances from acoustic sources
perturbations of $B$ represent acoustic waves (Howe, 2001).

The terms on the right hand side of Eq. (\ref{eq:215}) represent
acoustic sources:
\begin{equation}
S_L \equiv \left( {\bf \nabla} + \frac{{\bf \nabla}p}{\rho c_s^2}
\right) \cdot \left( {\bf \omega} \times {\bf v} \right),
\label{eq:216}
\end{equation}
\begin{equation}
S_T \equiv -\left( {\bf \nabla} + \frac{{\bf \nabla}p}{\rho c_s^2}
\right) \cdot \left( T{\bf \nabla}s \right), \label{eq:217}
\end{equation}
\begin{equation}
S_q \equiv \frac{\partial }{\partial t} \left( \frac{\gamma
T}{c_s^2} \frac{Ds}{Dt} \right) + \left( {\bf v} \cdot {\bf
\nabla}\right) \left( \frac{T}{c_s^2} \frac{Ds}{Dt} \right),
\label{eq:218}
\end{equation}
\begin{equation}
S_m \equiv \frac{\partial }{\partial t} \left( \frac{a}{1+aq}
\frac{Dq}{Dt} \right), \label{eq:219}
\end{equation}
\begin{equation}
S_\gamma \equiv p\frac{\partial \gamma}{\partial q} \left(
\frac{\partial q}{\partial t} \left({\bf v} {\bf \nabla} \right) p -
\frac{\partial p}{\partial t} \left({\bf v} {\bf \nabla} \right) q
\right). \label{eq:220}
\end{equation}

To make further mathematical analysis more tractable we will make
standard simplifying assumptions: for low Mach number flow all
convective derivatives in Eq. (\ref{eq:215}) can be replaced by time
derivatives $\partial /
\partial t$ (Goldstein, 1976); for the acoustic pressure in far
field we can use
\begin{equation}
p^\prime({\bf x},t) \approx \rho_0 B({\bf x},t); \label{eq:32}
\end{equation}
we neglect nonlinear effects of acoustic wave propagation and
scattering of sound by vorticity; we also neglect the influence of
stratification on the acoustic wave generation process and consider
background thermodynamic parameters in Eq. (\ref{eq:215}) as
constants. The later assumption is valid if the wavelength of
generated acoustic waves do not exceed the stratification length
scale (Stein, 1967),
\begin{equation}
\lambda \lesssim H \equiv \frac{c_s^2}{g} \approx 10^4 {\rm m}.
\label{eq:31}
\end{equation}

With these assumptions Eq. (\ref{eq:215}) reduces to
\begin{eqnarray}
\frac{1}{\rho_0} \left( \frac{1}{c_s^2} \frac{\partial^2}{\partial
t^2} - {\bf \nabla}^2 \right) p^\prime = S_L + S_T + S_\gamma + S_q
+ S_m \label{eq:33}
\end{eqnarray}
with
\begin{equation}
S_L \approx {\bf \nabla} \cdot \left( {\bf \omega} \times {\bf v}
\right), \label{eq:34}
\end{equation}
\begin{equation}
S_T \approx -{\bf \nabla} \cdot \left( T{\bf \nabla}s \right),
\label{eq:35}
\end{equation}
\begin{equation}
S_\gamma = p\frac{\partial \gamma}{\partial q} \left( \frac{\partial
q}{\partial t} \left({\bf v} {\bf \nabla} \right) p - \frac{\partial
p}{\partial t} \left({\bf v} {\bf \nabla} \right) q \right).
\label{eq:38}
\end{equation}
\begin{equation}
S_m \approx \frac{a}{1+aq} \frac{\partial^2q}{\partial t^2},
\label{eq:37}
\end{equation}
\begin{equation}
S_q \approx -\frac{\gamma L_\nu}{c_s^2} \frac{\partial^2q}{\partial
t^2}, \label{eq:36}
\end{equation}

First three terms on the right hand side of Eq. (\ref{eq:33})
represent well known sources of sound: the first term represents
Lighthill's quadrupole source (Lighthill, 1952); the second term is
a dipole source related to temperature fluctuations (Goldstein,
1976); $S_\gamma$ is a monopole source related to variability of
adiabatic index and usually has negligible acoustic output (Howe,
2001). As it was shown in Akhalkatsi and Gogoberidze (2009), in the
case of saturated moist air turbulence there exists two additional
monopole sources of sound, $S_q$ and $S_m$, related to nonstationary
heat and mass production during condensation of moisture,
respectively.

$S_q$ and $S_m$ produce monopole radiation and physically have the
following nature: suppose there exist two saturated air parcels of
unit mass with different temperatures $T_1$ and $T_2$ and water
masses $m_\nu(T_1)$ and $m_\nu(T_2)$. Mixing of these parcels leads
to the condensation of water due to the fact that
\begin{equation}
2m_\nu(T_1/2+T_2/2) <m_\nu(T_1)+m_\nu(T_2). \label{eq:481}
\end{equation}

Condensation of water leads to two effects important for sound
generation: production of heat and decrease of gas mass. Both of
these effects are known to produce monopole radiation (Goldstein,
1976; Howe, 2001). Consequently, turbulent mixing of saturated air
with different temperatures will lead not only to dipole
thermo-acoustical radiation, but also to monopole radiation.

As it was shown by Akhalkatsi and Gogoberidze (2009), for typical
parameters of supercell storms acoustic radiation is dominated by a
monopole source related to the heat production $S_q$. The total
acoustic power can be estimated as
\begin{equation}
N_q \sim {\frac {\rho_0 {\gamma}^2 L^2_\nu \Delta q^2
M_t^4}{lc_s}}F_1, \label{eq:49}
\end{equation}
where $\Delta q$ is the rms of humidity mixing ratio perturbations;
$l$ is the length scale of energy containing turbulent eddies;
$M_t\equiv v/c_s\ll 1$ is turbulent Mach number and $F_1$ is the
volume occupied by saturated moist air turbulence. For typical
parameters $S_q$ is much greater, than other sources of sound and
two orders of magnitude stronger then Lighthill's quadrupole source.

\section{Spectral Decomposition} \label{sec:3}

In this section we study the spectrum of acoustic radiation related
to $S_q$. Dropping all other source terms (\ref{eq:33}) reduces to
the inhomogeneous wave equation
\begin{eqnarray}
\frac{1}{\rho_0} \left( \frac{1}{c_s^2} \frac{\partial^2}{\partial
t^2} - {\bf \nabla}^2 \right) p^\prime = S_q. \label{eq:39}
\end{eqnarray}

Using standard methods for spectral analysis of the inhomogeneous
wave equation (Goldstein, 1976; Howe, 2001; Gogoberidze \emph {et
al}., 2007), after a long but straightforward calculation, for the
spectrum of temperature fluctuations $I({\bf x},\omega)$ we obtain
\begin{equation}
I({\bf x},\omega) =\frac{\omega^4 \pi\rho_0\gamma^2 L^2_\nu}{2
c_s^5|{\bf x}|^2} {\left(\frac{\partial^2 q_s}{\partial
T^2}\right)}^2 \int {\rm d}^3 {\bf x}^\prime H \left( {\bf
x}^\prime, \frac{{\bf x}}{|{\bf x}|} \omega, \omega\right),
\label{eq:417}
\end{equation}
where $H({\bf x}^\prime,{\bf k},\omega)$ is a spectral tensor of
temperature fluctuations and represents a Fourier transform of a two
point time delayed forth order correlation function of temperature
fluctuation.

Equation (\ref{eq:417}) allows to calculate the spectrum of sound
radiated by a monopole source related to the moisture, if
statistical properties of the turbulent source can be determined.
For our calculations we consider the Von Karman model of stationary
and homogeneous turbulence, which is given by (Hinze, 1975)
\begin{equation}
E_k=C_K \varepsilon^{2/3} k_0^{-5/3} \frac{(k/k_0)^4}{\left[
1+(k/k_0)^2 \right]^{17/6}}.
 \label{eq:419}
\end{equation}
for $k<k_d$ and $E_k=0$ for $k>k_d$, where ${2\pi}/{k_d}$ is the
dissipation length scale.

The Von Karman spectrum reduces to the Kolmogorov spectrum in the
inertial interval ($k\gtrsim k_0$), but also satisfactorily
describes the energy spectrum in the energy containing interval,
which is a dominant contributor to acoustic radiation.

As is known (Monin and Yaglom, 1975), temperature fluctuations of
homogeneous isotropic and stationary turbulence behaves like a
passive scalar and therefore has the same spectrum as velocity
fluctuation. Therefore for the spectral function of temperature
fluctuation $F({\bf k},\tau)$, which is spatial Fourier transform of
temperature correlation function $\Theta(\bf r,t)=\left\langle
T^\prime(\bf {x+r},t)T^\prime (\bf {x},t)\right\rangle$, we assume
\begin{equation}
F({\bf k},\tau)= \frac{Q_k}{4\pi k^2} f(\eta_k,\tau), \label{eq:418}
\end{equation}
where
\begin{equation}
Q_k=\Delta T^2 k_0^{-1} \frac{(k/k_0)^4}{\left[ 1+(k/k_0)^2
\right]^{17/6}}.
 \label{eq:420}
\end{equation}
In equation Eq. (\ref{eq:418}) $\triangle T$ is rms of the
temperature fluctuation, $\eta_k$ is the autocorrelation function
defined as
\begin{equation}
\eta_k=\sqrt{\frac{k^3 E_k}{2\pi}}
 \label{eq:421}
\end{equation}
and $f(\eta_k,\tau)$ characterizes the temporal decorrelation of
turbulent fluctuations, such that it becomes negligibly small for
$\tau \gg 1/\eta_k$.

For $f(\eta_k,\tau)$ we use a square exponential time dependence
(Kraichnan, 1964)
\begin{equation}
f(\eta_k,\tau)=\exp \left( -\frac{\pi}{4} \eta^2_k \tau^2 \right).
\label{eq:425}
\end{equation}

For a homogeneous turbulence two point time delayed forth order
correlation function of temperature fluctuations $R({\bf x}^\prime,
{\bf x}^\prime + {\bf r}, \tau)$ is related to the temperature
correlation function by means of the following relation (Monin and
Yaglom, 1975)
\begin{equation}
R({\bf x}^\prime, {\bf x}^\prime + {\bf r}, \tau) = 2\left\langle
T^\prime({\bf r}, \tau)T^\prime({\bf r},
\tau)\right\rangle\left\langle T^\prime({\bf r}, \tau)T^\prime({\bf
r}, \tau)\right\rangle. \label{eq:427}
\end{equation}
Using Eqs. (\ref{eq:418}),(\ref{eq:421}),(\ref{eq:425}) and the
convolution theorem we find
\begin{equation}
H({\bf k},\omega) = 2 \int {\rm d} {\bf k}^3_1 {\rm d} \omega_1
g({\bf k}_1,\omega_1) g({\bf k} - {\bf k}_1,\omega - \omega_1),
\label{eq:428}
\end{equation}
where
\begin{equation}
g({\bf k},\omega) = \frac{Q_k}{4\pi^2 \eta_k} \exp \left(-
\frac{\omega^2}{\pi \eta^2_k k^2} \right). \label{eq:429}
\end{equation}
For low Much number turbulence one can use the so called
aeroacoustic approximation (Goldstein, 1976), which for the Fourier
spectrum implies that in Eq. (\ref{eq:417}) instead of $H \left(
{\bf x}^\prime, \omega {\bf x}/|{\bf x}|, \omega\right)$ one can set
$H \left( {\bf x}^\prime, 0, \omega\right)$.

Performing the integration with respect to frequency in Eq.
(\ref{eq:428}) as well as angular variables in wave number space we
obtain
\begin{equation}
H(0,\omega) = \frac{\sqrt{2}}{4\pi^2} \int_{k_0}^{k_d} {\rm d}k
\frac{Q_k^2}{k^2\eta_k} \exp\left( -\frac{\omega^2}{2\pi \eta_k^2}
\right). \label{eq:430}
\end{equation}

The aeroacoustic approximation simplifies finding asymptotic limits
for the spectrum. In the low-frequency regime, taking the limit
$\omega\rightarrow 0$ we obtain
\begin{equation}
H (0,\omega)\sim \frac {1}{10\pi^{3/2}} \frac{\Delta T^4}{k^4_0
Mc_s}. \label{eq:431}
\end{equation}
For the spectrum this means $I({\bf x},\omega)\backsim \omega^4$.
Physically, these frequencies are lower than the lowest
characteristic frequency in the problem, corresponding to the eddy
turnover time on the energy containing scale.

At high frequencies $\omega \gg k_0 MR^{1/2}$, the integral in Eq.
(\ref{eq:430}) is dominated by the contribution from its upper limit
and we get
\begin{equation}
H (0,\omega)\sim \frac {3}{8\pi^{3/2}} \frac{\Delta T^4}{k^2_0}
\frac{Mc_s}{\omega ^2} \exp\left(-\frac
{\omega^2}{k_0^2M^2c_s^2R}\right) \label{eq:432}
\end{equation}
and
\begin{equation}
I({\bf x},\omega)\sim \omega^2 \exp\left(-\frac
{\omega^2}{k_0^2M^2c_s^2R}\right). \label{eq:432a}
\end{equation}
The functional form of the high-frequency suppression is determined
by the specific form of the time autocorrelation function of
turbulence Eq. (\ref{eq:425}) (Kraichnan, 1964), but for any
autocorrelation the amplitude of emitted waves should be very small
in this band. Physically, this limit corresponds to radiation
frequencies which are larger than any frequencies of turbulent
motions; consequently, no scale of turbulent fluctuations generates
these radiation frequencies directly, and the resulting small
radiation amplitude is due to the sum of small contributions from
many lower-frequency source modes. Since the integral is dominated
by the upper integration limit, the highest-frequency source
fluctuations (which contain very little of the total turbulent
energy) contribute most to this high-frequency radiation tail.

In the intermediate frequency regime, $k_0Mc_s < \omega <
k_0Mc_sR^{1/2}$, the integral  in Eq. (\ref{eq:430}) is dominated by
the contribution around $k_1$, where $\eta_{k_1}\approx \omega$. The
width of the dominant interval is $\Delta k_1\sim k_1$. Physically,
this implies that radiation emission at some frequency in this range
is dominated by turbulent vortices of the same frequency.
Consequently, for the inertial interval we obtain following estimate
\begin{equation}
H(0,\omega)\simeq\frac{1}{3\pi^{3/2}}\frac{\Delta T^4}{k^4_0
Mc_s}\left(\frac{k_0 M c_s}{\omega}\right)^{15/2} \label{eq:433}
\end{equation}
and
\begin{equation}
I({\bf x},\omega)\sim \omega^{-7/2}. \label{eq:433a}
\end{equation}

We performed numerical integration of Eq. (\ref{eq:430}) for the Von
Karman model of turbulence and determined normalized spectrum of
acoustic radiation to be
\begin{equation}
I_N(\nu)=\frac{4\sqrt{2}\pi c_s^2 {|\bf x|}^2}{\rho_0 \gamma^2
L_\nu^2 k_0 v_0^4 \Delta T^4 F_1}I({\bf x},\omega).
\end{equation}
The normalized spectrum for characteristic length scale of energy
containing eddies $l=2\pi/k_0=15~{\rm m}$ and characteristic rms
velocity $v_0=5~{\rm m~s^{-1}}$ is presented in Fig. 1. As can be
seen for these typical parameters the peak frequency of infrasound
radiation is $\nu_{peak} \approx 0.8~{\rm Hz}$.

As shown by Akhalkatsi and Gogoberidze (2009), the peak frequency of
 acoustic radiation is inversely proportional to the turnover time
of energy containing turbulent eddies $\nu_{peak} \sim v_0/l$,
whereas total acoustic power is proportional to $v_0^4$, $\Delta
T^4$ and inversely proportional to $l$.

\section{Infrasound correlation with tornadoes} \label{sec:6}

Severe storm forecasting operations are based on several large scale
environmental, storm scale, meso-beta scale kinematic and
thermodynamic parameters (Davies-Jones, 1993; Lemon and Doswell,
1979; Markowski \emph{et al}., 1998, Markowski \emph{et al}., 2002),
used to study the potential of severe weather, thunderstorm
structure and organization to produce tornadoes. Recent
climatological studies of thunderstorms using real-time radar data
combined with observations of near-storm environment have been
focused on the utility of various supercell and tornado forecast
parameters (CAPE, Storm Relative Helicity - SRH, Bulk Richardson
number - BRN and other parameters). Two parameters have been
established to be the most promising in discriminating between
nontornadic and tornadic supercells: boundary layer water vapor
concentration (LCL hight) and low level vertical wind shear
(Markowski and Richardson, 2008; Rasmussen, 2003; Thompson and
Edwards, 2000).

Examining a baseline climatology of parameters commonly used in
supercell thunderstorm forecasting and research, Rasmussen and
Blanchard (1998) established that the parameter that shows the most
utility for discriminating between soundings of supercells with
significant tornadoes and supercells without significant tornadoes
is LCL height. The height of the LCL appeared to be generally lower
for supercells with significant tornadoes than those without.
Rasmussen and Blanchard (1998) also found that for storms producing
large (at least 5-cm wide) hail only, without at least F2 strength
tornadoes, the LCL heights were significantly higher than for
ordinary thunderstorms. In their study half of the tornadic
supercells soundings had LCLs below 800 m, while LCL heights above
about 1200 m were associated with decreasing likelihood of
significant tornadoes. They concluded that stronger evaporational
cooling of moist downdraft leads to greater outflow dominance of
storms in high LCL settings and low heights increase the likelihood
of supercells being tornadic.

The work of Thompson and Edwards (2000) on assessing utility of
various supercell and tornado forecasting parameters supports the
finding that supercells above deeply mixed convective boundary
layers, with relatively large dew point depressions and high LCL,
often do not produce tornadoes even in environments of large CAPE
and/or vertical shear. They found the LCL to be markedly lower for
supercells producing significant tornadoes than for those producing
weak tornadoes, which were in turn lower than for nontornadic
supercells. Particularly, no strong and violent tornadoes occurred
for supercells with LCL $>1500$ m.

Studying the relationship between Rear flank downdraft (RFD)
thermodynamic characteristics and tornado likelihood Markowski
\emph{et al}. (2002) found that low LCL favors formation of
significant tornadoes because the boundary layer relative humidity
somehow alters the RFD and outflow character of supercells.
Relatively warm and buoyant RFDs, which are supposed to be necessary
for the genesis of significant tornadoes, were more likely in moist
low-level environments than in dry low-level environments (Markowski
\emph{et al}. 2002). It appeared that relatively dry boundary
layers, characterized by higher LCLs, support more low-level cooling
through the evaporation of rain, leading to stronger outflow, which
could have been decreasing the likelihood of significant tornadoes
in supercells. These are possible explanations for finding that the
LCL height is generally lower in soundings associated with tornadic
suppercells versus nontornadic (Rasmussen and Blanchard, 1998;
Thompson and Edwards, 2000).

Thompson \emph{et al}. (2003) reinforced the findings of previous
studies related to LCL height as an important discriminator between
significantly tornadic and nontornadic supercells and concluded that
the differences in LCL heights across all storm groups studied were
statistically significant, though the differences appeared to be
operationally useful only when comparing significantly tornadic and
nontornadic supercells. The lower LCL heights of the significantly
tornadic storms supported the hypothesis of Markowski \emph{et al}.
(2002) that increased low-level relative humidity (RH) may
contribute to increased buoyancy in the rear flank downdraft and an
increased probability of tornadoes.

In idealized numerical simulations Markowski \emph{et al}. (2003)
investigated the effects of ambient LCL and the precipitation
character of a rain curtain on the thermodynamic properties of
downdraft, and ultimately on tornado intensity and longevity. The
simulations were consistent with the observation that high boundary
layer relative humidity values (i.e., low LCL height and small
surface dewpoint depression) are associated with relatively warmer
RFDs and more significant tornadogenesis than environments of
relatively low boundary layer relative humidity.

These findings of low LCL favoring significant tornadoes could
explain observed high correlation between low frequency infrasound
signals from supercell storm and later tornado formation.

Acoustic power of a monopole source related to the heat production
during condensation of moisture can been estimated as (Akhalkatsi
and Gogoberidze, 2009):

\begin{equation}
N_q \sim {\frac {\rho_0}{lc_s}}{\left[\gamma L_\nu\right]}^2 {\left[
{\frac {M_t \Delta T}{T}}\right]}^4 f^2(T_c)F_1, \label{eq:501}
\end{equation}
where
\begin{equation}
f(T_c)\approx 6.8\cdot
10^4\frac{273.15+T_c}{(243.5+T_c)^4}\exp\left( \frac{17.67
T_c}{T_c+243.5}\right). \label{eq:502}
\end{equation}
and $T_c=T-273.15$ is the temperature in degree Celsius.

Due to the numerator in the exponent, $f(T_c)$ strongly depends on
temperature, e.g., $f(T_C=10^\circ)/f(T_C=0^\circ)\approx 2$ and
according to multiplier $f^2(T_c)$ in Eq. (\ref{eq:501}), increase
of saturated air temperature causes rapid enhancement of total
acoustic power radiated by a monopole source.

On the other hand, low LCL height means low level air in the updraft
motion being saturated and consequently, higher temperature of
saturated moist air. Therefore, the lower LCL heights contribute to
increased total acoustic power radiated by a monopole source related
to the heat production during the condensation of moisture. As a
result, enhanced low frequency infrasound signals from supercell
storm appear to be in strong correlation with an increased
probability of tornadoes.

It is also known (Weisman and Klemp, 1982; Rotunno and Klemp, 1982;
Rasmussen, 2003; Rasmussen and Blanchard, 1998; Thompson \emph {et
al}., 2003)that high values of supercell CAPE assist tornado
formation. Indeed, high values of CAPE lead to an increase of the
updraft persistence and thunderstorm activity and therefore increase
probability of tornado formation. According to recent studies
relatively larger CAPE occurs closer to the surface, which could
produce more intense low-level stretching of vertical vorticity
required for low-level mesocyclone intensification and perhaps
tornadogenesis (Rasmussen, 2003, McCaul, 1991; McCaul and Weisman,
1996) is associated with low LCLs. Rasmussen (2003) found the 0–3-km
above ground level (AGL) CAPE to be possibly important in
discriminating between environments supportive of significant
tornadoes and those that are not.

High values of CAPE mean high updraft velocity caused by large
low-level accelerations and increased rms of turbulent velocities.
According to Eq. (\ref{eq:501}) $N_q \sim M_t^4$, Therefore
increasing rms of turbulent velocities results in strong enhancement
of total acoustic power.

\section{Conclusions} \label{sec:7}

In this paper we have performed detailed spectral analysis of a
monopole source of sound related to heat production during
condensation of moisture, which is supposed to be dominant in the
infrasound radiation observed from strong convective storms. We have
also discussed the relationship between the acoustic power of this
source and certain significant tornado forecast parameters.
Particularly, low LCL, which is known to favor significant tornadoes
(Rasmussen and Blanchard, 1998; Thompson \emph {et al}., 2003)
implying warmer air at the level of saturation. We have shown that
the increase of temperature causes rapid enhancement of acoustic
power. High values of CAPE (especially, occurring closer to the
surface), which assist tornado formation (Weisman and Klemp, 1982;
Rotunno and Klemp, 1982; Rasmussen, 2003; Rasmussen and Blanchard,
1998; Thompson \emph {et al}., 2003), means high updraft velocity
and therefore, increased rms of turbulent velocities, which results
in strong enhancement of total acoustic power.

ISNeT data combined with the information from Doppler Radar may help
to improve tornado forecast and reduce false alarms from
non-tornadic supercells. Recent studies comparing ISNeT output with
occurrences of tornadoes (Bedard \emph {et al}., 2004a) and
correlating ISNeT signals with detailed radar output (Szoke \emph
{et al}., 2004) show, that infrasound of a tornadic thunderstorm is
much stronger than the infrasound of a nonsevere weather system.
Correlation between the intensity of infrasound signals and
probability of later tornado formation indicates the potential for
discriminating between supercells that produce tornadoes and those
that do not. Therefore, information from an infrasound detecting
system may help to determine potentially tornadic storms.

\acks G. G. acknowledges partial support from Georgian NSF
ST06/4-096 and 07/406/4-300 grants.


\newpage

\newpage

\begin{figure}
\centering
\includegraphics[width=5cm]{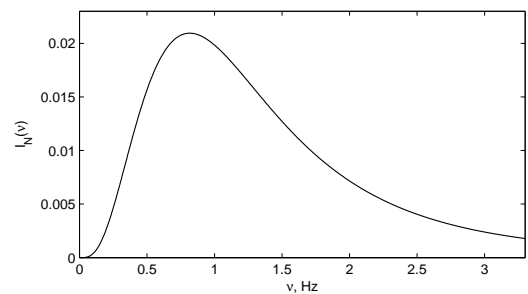}
\caption{Normalized spectrum od acoustic radiation for the Von
Karman turbulence.}
\end{figure}

\end{document}